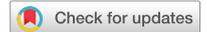

# Lumbar spine segmentation in MR images: a dataset and a public benchmark

Jasper W. van der Graaf[1,2] ✉, Miranda L. van Hooff[2,3], Constantinus F. M. Buckens[4], Matthieu Rutten[1,5], Job L. C. van Susante[6], Robert Jan Kroeze[7], Marinus de Kleuver[2], Bram van Ginneken[1] & Nikolas Lessmann[1]

This paper presents a large publicly available multi-center lumbar spine magnetic resonance imaging (MRI) dataset with reference segmentations of vertebrae, intervertebral discs (IVDs), and spinal canal. The dataset includes 447 sagittal T1 and T2 MRI series from 218 patients with a history of low back pain and was collected from four different hospitals. An iterative data annotation approach was used by training a segmentation algorithm on a small part of the dataset, enabling semi-automatic segmentation of the remaining images. The algorithm provided an initial segmentation, which was subsequently reviewed, manually corrected, and added to the training data. We provide reference performance values for this baseline algorithm and nnU-Net, which performed comparably. Performance values were computed on a sequestered set of 39 studies with 97 series, which were additionally used to set up a continuous segmentation challenge that allows for a fair comparison of different segmentation algorithms. This study may encourage wider collaboration in the field of spine segmentation and improve the diagnostic value of lumbar spine MRI.

## Background & Summary

Low back pain (LBP) causes the largest burden of disease worldwide, with most years lived with disability of any disease[1]. As a consequence, lumbar spine magnetic resonance imaging (MRI) for LBP is one of the most used imaging procedures within musculoskeletal imaging[2]. In the United States, 93% of the lumbar MRI referrals were appropriate according to the American College of Radiology guidelines, even though only 13% of the scans contributed in the clinical decision making[3]. Automatic image analysis might be the key to improve the diagnostic value of MRI by enabling more objective and quantitative image interpretation. A first step toward automatic assessment of lumbar spine MRI is segmentation of relevant anatomical structures, such as the vertebrae, intervertebral discs (IVDs) and the spinal canal.

With recent advances in machine learning and artificial intelligence (AI), state-of-the-art spine segmentation algorithms are generally learning-based algorithms that require well-curated training data. The development of vertebra segmentation algorithms for CT images has considerably benefitted from multiple large publicly available datasets with CT images and reference segmentations[4,5]. Currently no comparable large high quality datasets are available for lumbar spine MRI. Existing available datasets are either small, only segment the vertebral body[6,7], or are only annotated in the midsagittal slice (2D)[8,9]. Moreover, most datasets are limited to only one of the many anatomical structures that are most relevant for assessing multifactorial disorders such as LBP, i.e., only the vertebrae[10–14] or the IVDs[15–17].

To advance the development of segmentation algorithms, and ultimately automatic image analysis, for lumbar spine MRI, this study has three primary goals:

[1]Diagnostic Image Analysis Group, Radboud University Medical Center, Nijmegen, The Netherlands. [2]Department of Orthopedic surgery, Radboud University Medical Center, Nijmegen, The Netherlands. [3]Department Research, Sint Maartenskliniek, Nijmegen, The Netherlands. [4]Department of Medical Imaging, Radboud University Medical Center, Nijmegen, The Netherlands. [5]Department of Radiology, Jeroen Bosch Hospital, 's-Hertogenbosch, The Netherlands. [6]Department of Orthopedic Surgery, Rijnstate Hospital, Arnhem, the Netherlands. [7]Department of Orthopedic Surgery, Sint Maartenskliniek, Nijmegen, The Netherlands. ✉e-mail: jasper.vandergraaf@radboudumc.nl

| Hospital | Studies | T1 | T2 | T2 SPACE | Voxel size range (min – max)(mm) | Sex (% female) |
|---|---|---|---|---|---|---|
| UMC | 41 | 39 | 39 | 41 | (3.24 × 0.27 × 0.47) - (3.34 × 0.59 × 0.85)* | 55% |
| RH1 | 43 | 43 | 37 | 0 | (0.46 × 0.46 × 4.20) - (9.63 × 1.06 × 1.06) | 58% |
| RH2 | 44 | 24 | 44 | 0 | (0.46 × 0.46 × 4.20) - (5.17 × 1.00 × 1.23) | 59% |
| OH | 90 | 90 | 90 | 0 | (3.15 × 0.24 × 0.24) - (3.39 × 0.83 × 1.02) | 68% |
| Total | 218 | 196 | 210 | 41 | (3.15 × 0.24 × 0.24) - (9.63 × 1.06 × 1.23)* | 63% |

**Table 1.** Overview dataset. Abbreviations: UMC, University Medical Center; RH, Regional Hospital; OH, Orthopedic Hospital. *This only applies to the regular T1 and T2 weighted images. The T2 SPACE sequence has a voxel size of 0.90 × 0.47 × 0.47 mm.

1. To present a large multi-center lumbar spine MR dataset with reference segmentations of vertebrae, IVDs and spinal canal combined with per-level radiological gradings.
2. To introduce a continuous lumbar spine MRI segmentation challenge that allows algorithm developers to submit their models for evaluation.
3. To provide reference performance metrics for two algorithms that segment all three spinal structures automatically: a baseline AI algorithm, which was used in the data collection process, and the nnU-Net, a popular algorithm for 3D segmentation tasks for which training and inference code is publicly available.

## Methods

**Data collection.** In total, 257 lumbar spine studies from patients with a history of LBP were retrospectively collected, with each study consisting of up to three MRI series. Of these 39 patient studies, containing 97 MRI series, were sequestered for public benchmarking. The public data release described in this paper consists of 218 patient studies with 447 series. The study was approved by the institutional review board at Radboud University Medical Center (IRB 2016–2275). Informed consent was exempted, given the retrospective scientific use of deidentified MRI scans. Studies were collected from four different hospitals in the Netherlands, including one university medical center (UMC), two regional hospitals and one orthopedic hospital (data acquired between January 2019 and March 2022). All involved hospitals signed either a data transfer agreement or a public data sharing form in which public sharing of the data under a CC-BY 4.0 license was disclosed.

Data originating from the UMC were all available lumbar spine MRI studies of patients presenting with (chronic) low back pain between January 2019 and November 2020 that included a T2 SPACE sequence. This sequence produces images with almost isotropic spatial resolution (voxel size: 0.90 × 0.47 × 0.47 mm). All studies also contained both a standard sagittal T1 and T2 sequence (voxel size: 3.30 × 0.59 × 0.59 mm). MRI studies were only excluded if the image quality was considered too low for fully manual segmentation (n = 4). Data originating from the other three hospitals were sets of consecutive lumbar spine MRI studies of patients presenting with (chronic) low back pain with at least a sagittal T1 or a sagittal T2 sequence. The voxel size of these images ranged from 3.15 × 0.24 × 0.24 mm to 9.63 × 1.06 × 1.23 mm. At each center, we included a fixed number of consecutive MRI studies that met these inclusion criteria. There were no other exclusion criteria except for a number of studies across all four contributing centers being excluded from publication to serve as hidden hold-out test set for an algorithm-development challenge (see the Segmentation data section for further details). Additional dataset characteristics are given in Table 1.

*Segmentation data.* In all included MRI series, all visible vertebrae (excluding the sacrum), intervertebral discs, and the spinal canal were manually segmented. The segmentation was performed by a medical trainee who was trained and supervised by both a medical imaging expert (JG) and an experienced musculoskeletal radiologist (SB). Three-dimensional MRI annotation is a complex and laborious task, especially for the vertebral arch of the lumbar vertebrae. Therefore, we worked with an iterative data annotation approach in which our automatic baseline segmentation method (baseline 1: iterative instance segmentation) was trained with a small part of the dataset, enabling semi-automatic segmentation of the remaining images. During semi-automatic segmentation, the automatic method was used to obtain an initial segmentation, which was subsequently reviewed and manually corrected. This process was repeated several times by retraining the automatic segmentation model until the entire dataset was annotated.

Initially, twenty randomly selected high resolution T2 (SPACE) series of the UMC data were manually annotated using 3D Slicer version 5.0.3[18]. All structures were segmented in their entirety, which for the vertebrae also includes the vertebral arch. This was done since the vertebral arch is essential in the diagnosis of disorders such as foraminal stenosis, facet joint arthrosis, and spondylolysis. The initial manual annotations were performed only on high resolution series because the near-isotropic resolution enables detailed viewing in sagittal, axial and coronal directions. Annotations of the corresponding standard sagittal T1 and T2 images were obtained by resampling the T2 SPACE segmentations to the resolution of the T1 and T2 images. The resampled segmentations were reviewed for misalignment due to patient movement between the acquisitions and corrected if needed. Image windowing was adjusted dynamically in 3D slicer by the user to enhance the visibility of relevant structures. All twenty fully manually annotated MRI studies were segmented by a medical trainee and were reviewed by JG.

All other segmentations were created by first generating initial segmentations using the automatic segmentation method trained with all data for which annotations were available at that point. The annotated portion of the dataset increased iteratively by repeating this process four times while adding newly annotated data to the data used for training the automatic method that generated initial segmentations. The first time the algorithm was trained only on the twenty fully manually annotated studies to predict the segmentations of the remaining UMC data. The following three iterations were done with a retrained version of the network to generate segmentation for the data from each different hospital. The review on corrections of the predicted segmentations was manually done slice by slice in 3D Slicer by JG. The diagram in Fig. 1 shows the iterative segmentation pipeline. The main benefit of using this approach was that larger and easier structures do not require manual delineation, enabling to focus on smaller details and imperfections. The partial volume effect in the non-isotropic images was handled by viewing the annotations in all three directions and by using the smoothing functionalities in 3D Slicer when appropriate.

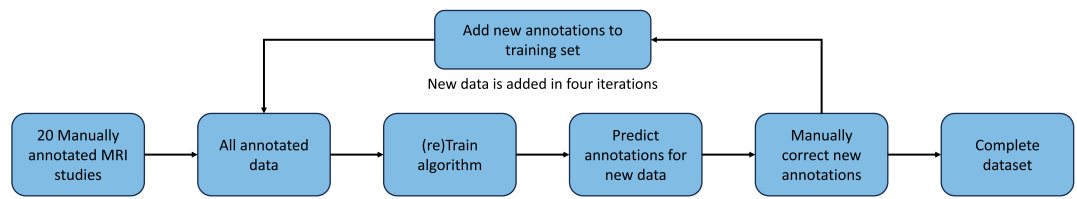

**Fig. 1** Diagram of the iterative annotation process.

All structures were given a separate segmentation label. The reference segmentations provided in this dataset are labelled from the bottom up with the most caudal lumbar vertebra labelled as 1. However, this label should not be interpreted as an anatomical label as this vertebra is not necessarily L5, the fifth lumbar vertebra, due to possible anatomical variations and irregular number of lumbar vertebrae. A lumbar-sacral transitional vertebra occurs in 35.5% of the population and the presence of a sixth lumbar vertebra in 6.6% of the population[19]. A larger field of view, axial MRI studies covering the complete lumbar spine or additional imaging are required to accurately determine the anatomical labels[20]. These were not available for the majority of studies in this dataset. The use of traditional clinical nomenclature was therefore considered infeasible due to the risk of mislabeling vertebrae.

For final evaluation of the used algorithm, all data (n = 257) was divided into a training set (n = 179), a validation set (n = 39), and a test set (n = 39). The test set was removed from the publicly available dataset to allow for public benchmarking (https://spider.grand-challenge.org/) and to avoid overfitting for fair comparison. This test set consists of 39 lumbar MRI studies of unique patients, which includes 15 out of the 20 fully manually annotated studies that were used in the iterative data annotation scheme. The remainder of the test set originates from the same four hospitals and was randomly selected in a similar distribution as the presented dataset. The 5 remaining fully manually annotated studies were placed in the validation set. The training and validation sets consist of 179 studies (82%) and 39 studies (18%) respectively. Series belonging to the same patient were always placed in the same set. The challenge is freely available to all users of the dataset, allowing other researchers to replicate the baseline results presented in this paper.

*Radiological gradings.* In order to allow users of this dataset to identify the more healthy and the more diseased cases in this dataset, the dataset includes radiological gradings at all IVD levels. Graded was either the presence or the severity of the following degenerative changes that can be observed on MRI images and that have a known or suspected relation to low back pain[21]:

1. Modic changes (type I, II or III)
2. Upper and lower endplate changes/Schmorl nodes (binary)
3. Spondylolisthesis (binary)
4. Disc herniation (binary)
5. Disc narrowing (binary)
6. Disc bulging (binary)
7. Pfirrman grade (grade 1 to 5)

All features were manually scored per IVD level by an expert musculoskeletal radiologist (MR, 26 years of experience).

**Baseline 1: Iterative instance segmentation.** By presenting this baseline algorithm, we establish a reference point for evaluating performance and provide users with an understanding of the algorithm employed in generating the dataset. This section summarizes the iterative instance segmentation (IIS) method. An automatic AI-based segmentation algorithm for vertebra segmentation[14] was extended to segment also the IVDs and the spinal canal. This algorithm uses a 3D patch-based iterative scheme to segment one pair of vertebra and the corresponding inferior IVD at a time, together with the segment of spinal canal covered by the image patch. A schematic image of the network architecture is shown in Fig. 2.

*Instance memory.* Because the MR volume is segmented by consecutively analyzing 3D patches, one vertebral level at a time, a method is needed to keep track of its progress. An instance memory volume is used to save the structures that have been segmented, and is used as an extra input channel to remind the network of the structures that can be ignored because they are already segmented. In contrast to the original vertebra-focused method, we introduced separate memory state volumes for the vertebrae, IVDs, and the spinal canal. The spinal canal memory state is only used to save the segmentation progress, not as an extra input for the network, as the spinal canal is an elongated structure that cannot be covered by a single patch. Therefore, the network is trained to always segment any visible portion of the spinal canal, which is then stitched together for all patches that are fed through the network. In total, the network has three input channels, two memory states, and the corresponding image patch.

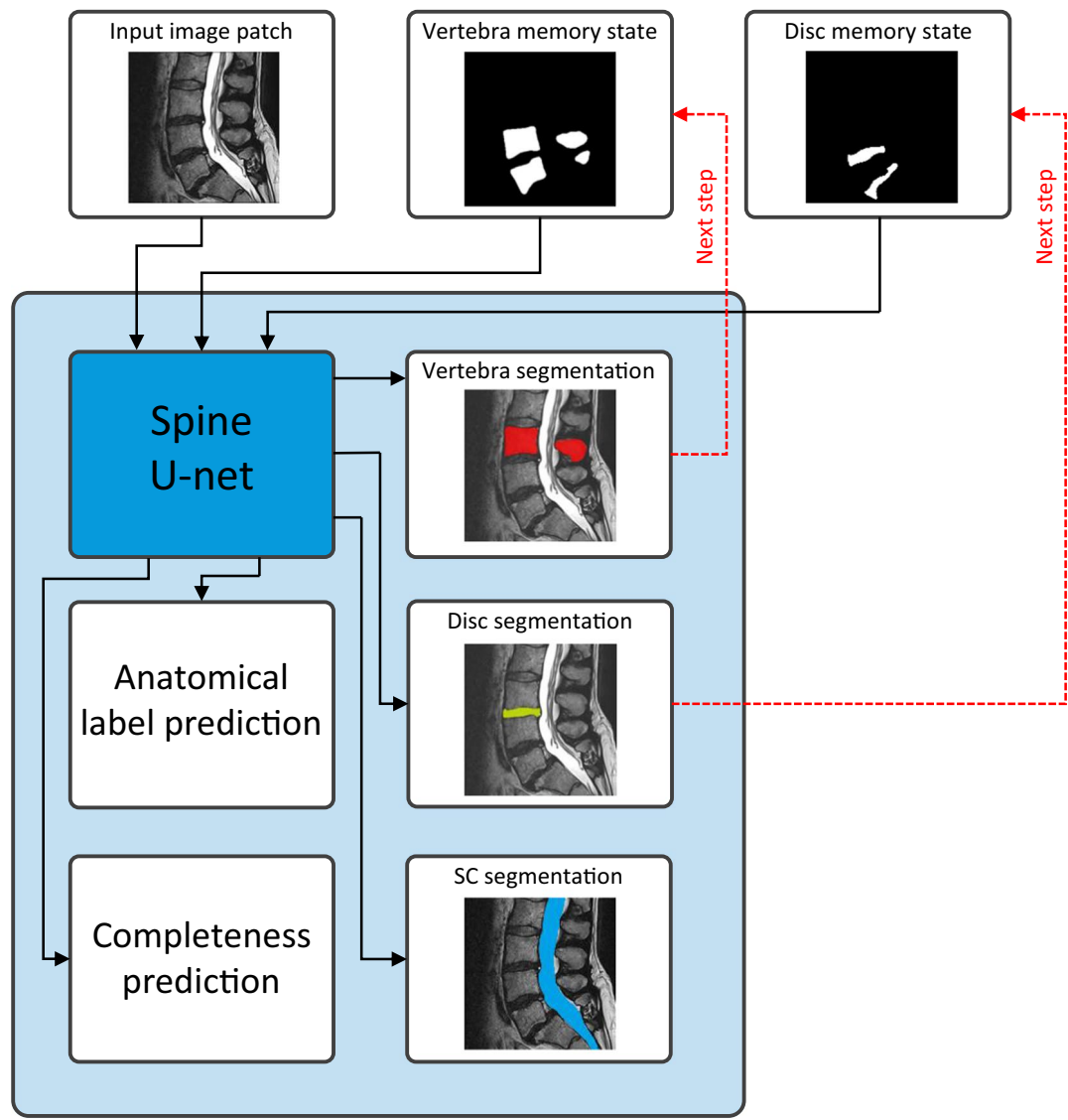

**Fig. 2** Schematic drawing of the network architecture. The input image and both memory states are fed into the Spine U-net, which produces predictions for the segmentation of the vertebrae (red), IVD (yellow), spinal canal (blue), as well as anatomical and completeness predictions. The vertebra and IVD segmentations are added to their respective memory states, which will be used as input to the network in the next iteration (see the "iterative segmentation approach" section for more detail).

*Network architecture.* The segmentation approach is based on a single 3D U-net-like fully-convolutional neural network. Unlike the vertebra segmentation algorithm as described in the original paper[14], a patch size of 64 × 192 × 192 voxels with a resolution of 2 × 0.6 × 0.6 mm was used, as the created dataset contains sagittal MR images exclusively. These generally have a higher slice thickness compared to the data used by Lessmann *et al.*[14]. A higher in-plane resolution of the predicted segmentation is achieved while still ensuring the patch is large enough that a vertebra completely fits within one patch. The network has three output channels, one for each anatomical structure.

*Iterative segmentation approach.* The patch-based scheme is structured in such a way that only relevant parts of the MR volume are being processed. The patch systematically moves through the image until it finds a fragment of the first vertebra, in this case always the lowest vertebra. Subsequently, the patch moves to the center of mass of that fragment after which a new segmentation is made. This process continues until the vertebra's volume stabilizes, which means that the detected vertebra is completely visible within the patch. Binary masks of that vertebra, its underlying IVD, and the spinal canal are then added to their respective memory states. The same patch is segmented again with the updated memory states as input, which causes a fragment of the next vertebra to be segmented. This iterative process, illustrated in Fig. 3, continues until no more vertebra fragments are detected or when the top of the MR volume is reached.

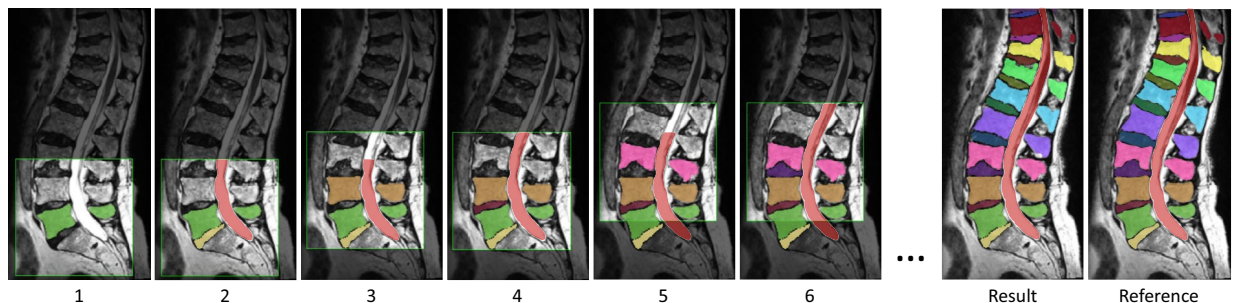

**Fig. 3** Illustration of the iterative segmentation approach's functionality. The process involves traversing the 3D patch (depicted by the light area with a green border) along the spine with alternating steps for segmentation of the vertebrae (shown in images 1, 3, and 5) and IVD and spinal canal (displayed in images 2, 4, and 6). The right-hand side displays the final automatic segmentation result alongside the reference segmentation.

*Completeness and label prediction.* The most cranial vertebra is often only partially visible within the field of view of the MR image. The segmentation method includes an additional compression path after the compression path of the U-net, which has a single binary value as output, predicting the completeness of a vertebra. The original vertebra segmentation method also contained a similar compression path for predicting the anatomical label. However, this output was not used in our experiments since no accurate anatomical labels regarding lumbosacral transitional vertebrae were present in our dataset.

*Training of the algorithm.* Preprocessing of the images consisted of resampling to a standard resolution of $2 \times 0.6 \times 0.6$ mm and orientation in axial slices. Standard data augmentation steps were implemented, such as random elastic deformation, the addition of random Gaussian noise, random Gaussian smoothing, and random cropping along the longitudinal axis. The loss function used during training consisted of three parts: (1) The segmentation error was defined by the weighted sum of false positives and false negatives combined with the binary cross-entropy loss. (2) The labeling error was defined by the absolute difference between the predicted label and the ground truth. (3) The completeness classification error was defined as the binary cross-entropy between the true label and the predicted probability. For final evaluation, the algorithm was trained using the training dataset while using the validation set to monitor the training process.

**Baseline 2: nnU-Net.** In addition to adapting a segmentation method that was specifically developed for vertebra segmentation, reference results for nnU-Net are provided. nnU-Net is a self-configuring, deep learning-based framework for medical image segmentation[22]. It has been widely accepted in the medical image analysis community as a state-of-the-art approach to 3D image segmentation tasks after winning the Medical Segmentation Decathlon[23] and performing well in several other segmentation challenges. A 3D full resolution nnU-Net was trained on the training and validation datasets with 5-fold cross validation, which is its recommended training strategy[22]. Data pre-processing, network architecture and other training details were automatically determined by the nnU-Net framework. The network was trained on both the T1- and T2-weighted MRI series after which the overall performance was compared to the IIS baseline algorithm.

**Evaluation.** The segmentation performance was evaluated using two metrics: (1) The Dice coefficient (measured in 3D) to measure the volume overlap, and (2) the average absolute surface distance (ASD) as an indication of the segmentation accuracy along the surface of all structures. Both metrics were calculated separately for all individual structures and were averaged per anatomical structure (vertebrae, IVDs, or spinal canal). Additionally, the average Dice coefficient and average ASD per MRI sequence (T1 vs. T2) were calculated for each anatomical structure. To ensure the Dice score and ASD are not influenced by labeling differences, the individual structures of the reference segmentation are matched to the structured in the predicted segmentation based on the largest found overlap. The completeness classification performance was determined by the percentage of accurate predictions, as well as the average number of false positives and false negatives. Evaluation was performed on a sequestered test set which is a subset of the presented dataset.

|  | Training | Validation | Total |
|---|---|---|---|
| Studies | 179 | 39 | 218 |
| Series | 360 | 87 | 447 |
| Vertebrae | 2512 | 613 | 3125 |
| IVD's | 2518 | 629 | 3147 |
| Sex (% female) | 63% | 62% | 63% |

**Table 2.** Overview of the distribution of data between the training and validation set. Abbreviation: IVDs, Intervertebral Discs.

|  | Training set | | Validation set | |
|---|---|---|---|---|
|  | IVDs (N = 1240) | Patients (N = 179) | IVDs (N = 280) | Patients (N = 39) |
| Modic I | 4 (0.3%) | 3 (1.7%) | 0 (0.0%) | 0 (0.0%) |
| Modic II | 376 (30.3%) | 115 (64.2%) | 127 (45.4%) | 31 (79.5%) |
| Modic III | 6 (0.5%) | 5 (2.8%) | 1 (0.4%) | 1 (2.6%) |
| Upper endplate changes | 472 (38.1%) | 133 (74.3%) | 142 (50.7%) | 34 (87.2%) |
| Lower endplate changes | 477 (38.5%) | 138 (77.1%) | 146 (52.1%) | 33 (84.6%) |
| Spondylolisthesis | 33 (2.7%) | 30 (16.8%) | 9 (3.2%) | 9 (23.1%) |
| Disc herniation | 62 (5.0%) | 55 (30.7%) | 10 (3.6%) | 10 (25.6%) |
| Disc narrowing | 440 (35.5%) | 158 (88.3%) | 104 (37.1%) | 35 (89.7%) |
| Disc bulging | 592 (47.5%) | 163 (91.1%) | 154 (55.0%) | 37 (94.9%) |
| Pfirrman 1 | 241 (19.4%) | 66 (36.9%) | 45 (16.1%) | 16 (41.0%) |
| Pfirrman 2 | 260 (21.0%) | 106 (59.2%) | 81 (28.9%) | 14 (35.9%) |
| Pfirrman 3 | 348 (28.1%) | 133 (74.3%) | 70 (25.0%) | 30 (76.9%) |
| Pfirrman 4 | 240 (19.4%) | 92 (51.4%-w | 51 (18.2%) | 25 (64.1%) |
| Pfirrman 5 | 151 (12.2%) | 79 (44.1%) | 51 (18.2%) | 16 (41.0%) |

**Table 3.** Overview of the radiological gradings per intervertebral disc (IVD) level and per patient for the training and validation set.

## Data Records

The complete dataset can be found at https://doi.org/10.5281/zenodo.10159290 and is available under the CC-BY 4.0 license[24]. Which MRI studies are assigned to the training and validation sets can be found in the overview file. This file also provides the biological sex for all patients and the age for the patients for which this was available. It also includes a number of scanner and acquisition parameters for each individual MRI study. The dataset also comes with radiological gradings found in a separate file. All radiological gradings are provided per IVD level.

To generate this dataset, a total of 218 lumbar MRI studies of patients with low back pain were included. Each study consisted of up to three sagittal MRI series which were either T1-weighted or T2-weighted (regular resolution, or high resolution generated using a SPACE sequence) with a total of 447 series. Of all included patients, 63% was female. A total of 3125 vertebrae, 3147 IVDs, and 447 spinal canal segmentations were included over all series combined. An overview of the complete dataset divided by the different hospitals is shown in Table 1. An overview of the training and validation sets and all included structures is shown in Table 2. The radiological gradings per IVD level and per patient are summarized in Table 3.

All MR images and their corresponding segmentation masks used in this study are stored in MHA format in separate directories. Both files have the same name, which is a combination of the MRI study identifier and the specific sequence type (T1, T2, or T2 SPACE). It is important to note that all MRI series from the same MRI study have the same identifier.

## Technical Validation

The performance of the IIS baseline algorithm, which was used to generate initial segmentation masks of unseen images from the dataset, was assessed on a hidden test set. These results are presented to assess the data-annotation strategy, as well as establish a reference performance for users of the dataset. The results for the different structures and the different sequences are shown in Table 4. The overall mean (SD) Dice score was 0.93 (±0.05), 0.85 (±0.10) and 0.92 (±0.04) for the vertebrae, IVDs and spinal canal respectively. The overall mean (SD) ASD was 0.49 mm (±0.95 mm), 0.53 mm (±0.46 mm) and 0.39 (mm ± 0.45) mm for the vertebrae, IVDs and spinal canal respectively. The spinal canal was identified in all scans. One of the 656 vertebrae and nine (three in T1 images and six in T2 images) of the 688 IVDs were not found. The completeness prediction was correct in 650 of the 656 vertebrae (99.1%). A nnU-Net was trained on the same training data to enable comparison between the IIS baseline algorithm and nnU-Net baseline. The results of both networks are displayed in Table 5. Figure 4 shows a collection of segmentations obtained by both networks.

|  |  | T1 | | T2 | | Total | |
|---|---|---|---|---|---|---|---|
|  |  | Mean | SD | Mean | SD | Mean | SD |
| Vertebrae | Dice | 0.93 | 0.04 | 0.92 | 0.06 | 0.93 | 0.05 |
|  | ASD (mm) | 0.45 | 0.84 | 0.51 | 1.01 | 0.48 | 0.95 |
|  | Detection% | 100.0% |  | 99.7% |  | 99.8% |  |
|  | Completeness accuracy% | 98.9% |  | 99.2% |  | 99.1% |  |
| IVDs | Dice | 0.85 | 0.10 | 0.84 | 0.10 | 0.84 | 0.10 |
|  | ASD (mm) | 0.48 | 0.47 | 0.58 | 0.43 | 0.54 | 0.45 |
|  | Detection% | 99.2% |  | 98.7% |  | 98.9% |  |
| Spinal canal | Dice | 0.93 | 0.03 | 0.92 | 0.04 | 0.92 | 0.04 |
|  | ASD (mm) | 0.29 | 0.17 | 0.46 | 0.56 | 0.39 | 0.45 |
|  | Detection% | 100.0% |  | 100.0% |  | 100.0% |  |

**Table 4.** Overview of all results of the IIS baseline algorithm. Abbreviations: IVDs, Intervertebral Discs; ASD, Average absolute Surface Distance; SD, Standard Deviation.

|  |  | IIS | | nnU-Net | |
|---|---|---|---|---|---|
|  |  | Mean | SD | Mean | SD |
| Vertebrae | Dice | 0.93 | 0.05 | 0.92 | 0.05 |
|  | ASD (mm) | 0.48 | 0.95 | 0.43 | 0.78 |
|  | Detection% | 99.8% |  | 100.0% |  |
| IVDs | Dice | 0.84 | 0.10 | 0.86 | 0.09 |
|  | ASD (mm) | 0.54 | 0.45 | 0.54 | 1.04 |
|  | Detection% | 98.7% |  | 100.0% |  |
| Spinal canal | Dice | 0.92 | 0.04 | 0.92 | 0.03 |
|  | ASD (mm) | 0.39 | 0.45 | 0.37 | 0.22 |
|  | Detection% | 100.0% |  | 100.0% |  |

**Table 5.** Comparison between the iterative segmentation algorithm and nnU-Net. Abbreviations: IVDs, Intervertebral Discs; IIS, Iterative Instance Segmentation; ASD, Average absolute Surface Distance; SD, Standard Deviation.

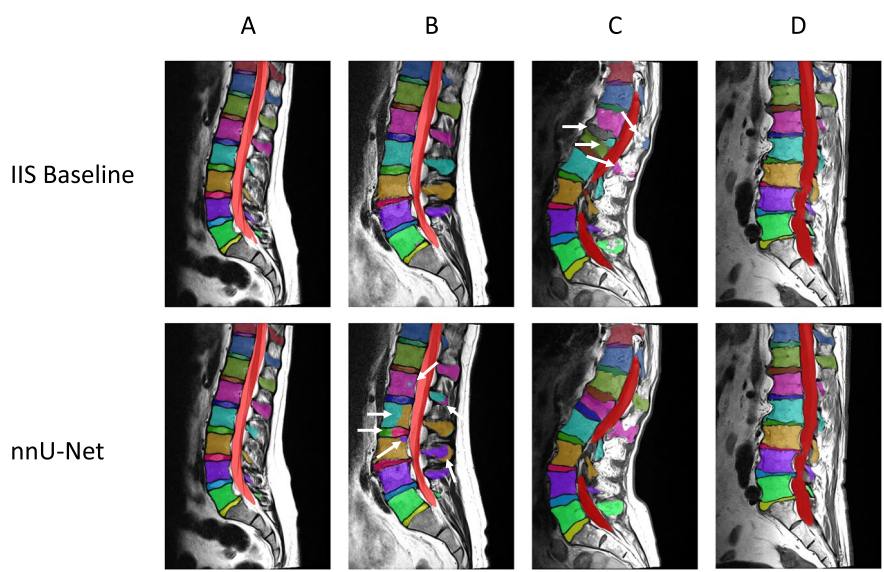

**Fig. 4** Examples of cases segmented by both baseline algorithms. Each column represents one case. Column (**A**) shows a case without any major pathologies and without significant segmentation errors. Columns (**B** and **C**) show cases where nnU-Net (**B**) and the IIS baseline algorithm (**C**) made mistakes, indicated by a white arrow. Column (**D**) shows a case with severe degenerative features present and no substantial segmentation errors.

The IIS model, which was used in the iterative annotation of the dataset, demonstrates strong performance on our dataset, which is comparable to other MR segmentation methods in the literature[7,25,26]. The performance of the IIS model was validated and benchmarked by comparing it against the performance of a second baseline algorithm. nnU-Net was chosen due to its widely acknowledged status as the current gold standard in medical image segmentation. The results of the two algorithms are nearly identical, indicating that the IIS baseline model was an accurate tool in the iterative data annotation workflow and is a reasonable benchmark for comparison.

The used iterative data annotation approach showed to be an effective strategy. One strength of this approach is its ability to improve the quality of the dataset over time by incorporating corrections from segmentation predictions into the training data. This helps to reduce errors and increase accuracy in subsequent iterations. Additionally, this approach was faster and more efficient compared to fully manual annotations. However, there are several limitations that should be addressed. Firstly, the iterative process of training the network on a small dataset, generating segmentation predictions on unseen images, and manually correcting the predictions before adding them to the dataset, can introduce bias in the final dataset. This strategy was chosen to shorten the time needed for manual annotation and thus enable the creation of a larger dataset. A key advantage of this approach was that larger and simpler areas did not need manual delineation. More intricate structures such as the vertebral lamina, the annulus of the IVDs, and epidural fat still regularly demanded manual corrections. On average, a fully manual segmentation took approximately five hours while correcting a predicted segmentation still took approximately one hour. Secondly, the use of only high-resolution T2 series for the initial manual annotation may not be representative of the entire population, as it is limited to patients from one hospital who underwent this specific imaging modality.

In the era of machine learning and AI algorithms, lumbar spine segmentation can serve as the basis for automated, accurate lumbar spine MR analysis, assisting clinical radiologists and imaging-minded spinal surgeons in their daily practice. It will be able to generate robust, quantitative MR results that can serve as inputs into larger models of lumbar spine disease in clinical practice and research settings. The availability of public datasets and benchmarks plays a crucial role in advancing the field. While datasets exist for CT vertebra segmentation, such as VerSe which is the largest available vertebra segmentation dataset[27], currently no public datasets for MRI spine segmentation are available. Our dataset is of similar size to VerSe[27] and provides full segmentation of all relevant spinal structures on MR images. This allows for wider participation and collaboration in the field of spine segmentation, as it can be used to train and evaluate algorithms, as well as to compare to other datasets. The presented algorithms are the baseline results to which other algorithms can be compared.

## Usage Notes

In order to allow for a fair comparison between different algorithms, including both baseline algorithms, a public segmentation challenge is hosted on the Grand Challenge platform and can be found at https://spider.grand-challenge.org/. Algorithms are evaluated on a sequestered test set which contains 97 MRI series of 39 unique patients.

## Code availability

The original training code of the IIS baseline algorithm and the trained weights and biases are publicly available at: https://github.com/DIAGNijmegen/SPIDER-Baseline-IIS. The nnU-Net baseline algorithm from Isensee *et al.* can be found here: https://github.com/MIC-DKFZ/nnUNet. Both trained algorithms can also be used on the grand challenge platform:

1. https://grand-challenge.org/algorithms/spider-baseline-iis/
2. https://grand-challenge.org/algorithms/spider-baseline-nnu-net

The source code used to compute the evaluation metrics on the test set are published at: https://github.com/DIAGNijmegen/SPIDER-Evaluation.

### Acknowledgements
This study was funded by Radboud AI for Health (ICAI).

### Author contributions
Jasper W. van der Graaf created the segmentation dataset, developed and trained the presented segmentation algorithms, and wrote the manuscript. Miranda L. van Hooff, Marinus de Kleuver, and Bram van Ginneken provided oversight for the project and revised the manuscript. Constantinus F.M. Buckens assisted with the MRI segmentation. Matthieu Rutten, Job van Susante, and Robert Jan Kroeze provided MRI data from their respective hospitals and revised the manuscript. Nikolas Lessmann helped with the development and training of the presented segmentation algorithms, provided overall project management, and revised the manuscript.

### Competing interests
The authors declare no competing interests.

### Additional information
**Correspondence** and requests for materials should be addressed to J.W.v.d.G.

**Reprints and permissions information** is available at www.nature.com/reprints.

**Publisher's note** Springer Nature remains neutral with regard to jurisdictional claims in published maps and institutional affiliations.